# Electron transport in DNA bases: An extension of the Geant4-DNA Monte Carlo toolkit

Sara A. Zein[1], Marie-Claude Bordage[2], Ziad Francis[3], Giovanni Macetti[4], Alessandro Genoni[4], Claude Dal Cappello[4], Wook-Geun Shin[1], Sebastien Incerti[1]

[1]Univ. Bordeaux, CNRS, CENBG, UMR 5797, F-33170 Gradignan, France

[2] Université Paul Sabatier, UMR1037 CRCT, INSERM, F-31037 Toulouse, France

[3]Saint Joseph University, Faculty of Sciences, U.R. Mathématiques et Modélisation, Beirut, Lebanon

[4] CNRS & Université de Lorraine, Laboratoire LPCT (UMR 7019), 1 Boulevard Arago, 57078 Metz, France

Corresponding Author: Sara A. Zein

Email addresses

zein@cenbg.in2p3.fr

marie-claude.bordage@inserm.fr

ziad.francis@gmail.com

giovanni.macetti@univ-lorraine.fr

Alessandro.Genoni@univ-lorraine.fr

claude.dal-cappello@univ-lorraine.fr

ukguen@gmail.com

incerti@cenbg.in2p3.fr

**Abstract**

The purpose of this work is to extend the Geant4-DNA Monte Carlo toolkit to include electron interactions with the four DNA bases using a set of cross sections recently implemented in Geant-DNA CPA100 models and available for liquid water. Electron interaction cross sections for elastic scattering, ionisation, and electronic excitation were calculated in the four DNA bases adenine, thymine, guanine and cytosine. The electron energy range is extended to include relativistic electrons. Elastic scattering cross sections were calculated using the independent atom model with amplitude derived from ELSEPA code. Relativistic Binary Encounter Bethe Vriens model was used to calculate ionisation cross sections. The electronic excitation cross sections calculations were based on the water cross sections following the same strategy used in CPA100 code. These were implemented within the Geant4-DNA option6 physics constructor to extend its capability of tracking electrons in DNA material in addition to liquid water. Since DNA nucleobases have different molecular structure than water it is important to perform more accurate simulations especially because DNA is considered the most radiosensitive structure in cells. Differential and integrated cross sections calculations were in good agreement with data from the literature for all DNA bases. Stopping power, range and inelastic mean free path calculations in the four DNA bases using this new extension of Geant4-DNA option6 are in good agreement with calculations done by other studies, especially for high energy electrons. Some deviations are shown at the low electron energy range, which could be attributed to the different interaction models. Comparison with water simulations shows obvious difference which emphasizes the need to include DNA bases cross sections in track structure codes for better estimation of radiation effects on biological material.

## 1. Introduction

Track structure Monte Carlo codes are computational tools used to simulate accurately ionizing radiation interactions within biological material, mainly approximated as liquid water [1]. They have many applications in radiobiology, medical physics and radioprotection [2]. Some of these codes, such as CPA100 [3] and KURBUC [4], not only simulate the physical stage but also the chemical stage of radiological electron-water interactions. An additional characteristic of CPA100 is the possibility to track electrons in DNA material [5]. PARTRAC code provides further options since it is able to simulate the different radiation interaction stages in water targets including DNA damage and repair [6]. The general purpose Monte Carlo code PENELOPE [7, 8] also provides detailed track structures for electrons down to 50 eV. Among them, the open source toolkit Geant4-DNA [9-12] simulates particle interactions step by step as a low energy physics extension of Geant4 [13-15] in liquid water. The physico-chemical and chemical radiation stages can also be simulated to estimate the yields of molecular species created from water radiolysis [16].

Since deoxyribonucleic acid (DNA) is considered the most radiation sensitive target within cells, accurate damage assessment is important. Previous studies approximate DNA damage through geometrical distribution of possible strand breaks within a homogeneous water medium [6, 17, 18]. DNA is a double stranded helical macromolecule forming the genetic code of living cells. It is composed of a sugar phosphate backbone and a long ladder-like sequence formed by four different nitrogenous bases: adenine, thymine, guanine and cytosine [19]. The integrity of this sequence is essential to the cell's health and replication and any damage or error in the sequence may result in carcinogenesis and cell death. A recent study by Francis *et al.* calculated the proton interaction cross sections within the four DNA bases using the semi-empirical Rudd model [20]. Total electron ionisation cross sections were also calculated with the BEB model. These cross sections were tested in Geant4-DNA where proton and electron stopping powers within the nucleobases were calculated in addition to lineal and specific energies of protons. Moreover, the current public version of Geant4-DNA provides interaction cross sections of electrons in DNA precursors tetrahydrofuran and trimethylphosphate ranging between 10 eV - 1 keV [21]. However, the electron interaction cross sections within DNA nucleobases are not yet available. Therefore, it is beneficial to perform direct determination of electron cross sections in DNA nucleobases and not derive them from DNA precursors.

Since 2017 Geant4-DNA has provided an alternative set of discrete physics models for the simulation of electron interactions in liquid water over the energies ranging between 11 eV and 256 keV under the "option6" physics constructor [12, 22]. These models implement the ionisation, electronic excitation and elastic scattering interactions of electrons obtained from the CPA100 track structure code developed by Terrissol *et al.* [3, 22]. In addition to physical interactions in liquid water, the original CPA100, coded in Fortran, also provides the water radiolysis simulation and radiation transport in different biological targets such as DNA bases [5]. For a better determination of radiation damage, it is important to consider a more realistic biological medium instead of using liquid water as the irradiated medium. The aim of this work is thus a continuation of our recently published work [22] in order to extend the "G4EmDNAPhysics_option6" constructor (which will be referred to as "option6" throughout the text and which contains CPA100 models for liquid water only) to include electron interactions in DNA material in addition to liquid water. Therefore, new sets of electron cross sections in the four DNA nucleobases, adenine ($C_5N_5H_5$), thymine ($C_5N_2O_2H_6$), cytosine ($C_4N_3OH_5$) and guanine ($C_5N_5OH_5$), were calculated specifically for this purpose. The new cross sections are extended to 1 MeV taking into account relativistic corrections of electron transport in the nucleobases, which are not DNA-bound as previously presented in the works of Edel [23] and Peudon [24].

In the following sections, the physics models for elastic scattering, electronic excitation and ionisation for electrons of energy range 11 eV – 1 MeV within the four nucleobases are presented. The interaction cross sections are calculated and compared to values from the literature. The physics models are implemented within Geant4-DNA according to the calculated cross sections. The stopping power, continuous-slowing-down approximation (CSDA) range and inelastic mean free path (IMFP) are calculated within the four nucleobases to verify the implementation and comparison with experimental and theoretical data are presented.

## 2. Materials and Methods
### 2.1 Physics models of electron interactions

In Geant4-DNA the new models for the physics processes of electrons in the four DNA bases (adenine, thymine, cytosine and guanine) are implemented on the basis of elastic scattering, electronic excitation and ionisation cross sections. The incident electron energy ranges from 11 eV to 1 MeV. The 11 eV lower limit is restricted by the elastic scattering, and the upper 1 MeV limit is chosen in accordance with the highest electron energy provided by the current Geant4-DNA version. The three different interactions cross sections governing the electron transport in the four DNA bases are calculated as described in the following sections.

#### 2.1.1 Elastic scattering

The elastic cross section in the four bases is calculated according to the well-known independent atom model (IAM) [25] similar to elastic cross section in water in the CPA100 code [22, 23]. In this approximation, the electron-molecule interaction is reduced to a collision with individual atoms constituting the molecule. This approach gives good results for many polyatomic molecules at high and intermediate incident energy corresponding to wavelengths shorter than the internuclear distances [25]. Therefore, the calculation of the elastic scattering differential cross section of a molecule requires the differential cross section of each atom $\frac{d\sigma_A}{d\Omega}$, the complex scattering amplitudes of the atoms $i$ ($f_i(\theta,k)$) and $j$ ($f_j^*(\theta,k)$), and the internuclear distance between the atoms $i$ and $j$ ($r_{ij}$).

$$\frac{d\sigma}{d\Omega} = \sum_{i=1}^{N}\frac{d\sigma_{A,i}}{d\Omega} + \sum_{i\neq j=1}^{N} f_i(\theta,k) f_j^*(\theta,k) \frac{\sin(sr_{ij})}{sr_{ij}} \qquad \text{Eq 1}$$

where $k$ is the incident electron wave number, $s$ ($s = 2k\sin\frac{\theta}{2}$) is the magnitude of the momentum transfer during the collision, $\theta$ is the scattering angle and $N$ is the total number of atoms in the target.

The geometry of the DNA bases was imported from the Chemical Structures Project, an open source software that provides 3D molecular structures for various molecules [26].

The scattering amplitudes are given by the ELastic Scattering of Electrons and Positrons by neutral Atoms (ELSEPA) code [27]. The elastic differential cross section per unit solid angle $d\Omega$ for atoms (H, C, N and O) are calculated using the corresponding scattering amplitudes derived from the partial wave expansion.

ELSEPA is a Fortran code developed by Salvat *et al.* [27]. This code allows not only to calculate electron elastic scattering differential and integrated cross sections but also to perform phase shift calculations for atoms with energies ranging from a few eV up to 1 GeV. Relativistic corrections are included in the Dirac partial wave approach within the static-exchange approximation. The interaction potential is the sum of different potentials (electrostatic, exchange, and correlation-polarization) and the Dirac partial wave analysis is performed. This code provides more recent and precise data compared to those that are used in CPA100, which is limited to low energy (<256 keV). Moreover, new elastic models in Geant4-DNA for gold [28] and water [29] targets were previously computed with ELSEPA.

### 2.1.2 Ionisation

The Binary Encounter Bethe (BEB) model for electron ionisation was initially developed by Kim and Rudd [30]. Since it has further developments and has been successfully used to calculate total cross sections and energy differential ones for a large number of atmospheric molecules [31], it was later used for industrial applications [32]. The agreement with experimental data is excellent for small size molecules at low energy. It has already been applied for water in CPA100 code [23] and "option6" within Geant4-DNA code [22].

The relativistic Binary Encounter Bethe Vriens model (RBEBV), developed by Guerra *et al.* [33], is used to calculate the ionisation cross section from the ionisation energy threshold to 1 MeV for each molecular orbital (MO) as a function of the incident kinetic energy T. This model also allows to calculate the energy of the ejected electron W using the energy differential cross section.

The analytical form of the cross section only depends on 3 parameters representative of the molecular orbital. This form is also convenient because it allows obtaining the energy loss by directly sampling this expression without using interpolation in large cross section tables [22].

The Monte Carlo track structure simulation codes require the precise knowledge of the energy of primary and ejected electrons after ionisation, which is given by the energy differential cross section (EDCS). The energy differential cross section for each molecular orbital $\frac{d\sigma_{ion,MO}}{dw}$ written in the reduced form is in the RBEBV model:

$$\frac{d\sigma_{ion,MO}}{dw} = \frac{4\pi a_0^2 \alpha^4 N}{(\beta_t^2 + \beta_u^2 + \beta_b^2)2b'}$$
$$\cdot \left[ -\frac{\phi}{t+1} \cdot \left(\frac{1}{w+1} + \frac{1}{t-w}\right) \cdot \frac{1+2t'}{(1+t'/2)^2} + \frac{1}{(w+1)^2} + \frac{1}{(t-w)^2} \right. \quad \text{Eq 2}$$
$$\left. + \frac{b'^2}{(1+t'/2)^2} + \left(Ln\left(\frac{\beta_t^2}{1-\beta_t^2}\right) - \beta_t^2 - Ln(2b')\right) \cdot \left(\frac{1}{(w+1)^3} + \frac{1}{(t-w)^3}\right) \right]$$

with $\quad w = \frac{W}{B}$

$\beta_t^2 = 1 - \frac{1}{(1+t')^2} \quad$ and $\quad t' = \frac{T}{mc^2} \quad$ and $\quad t = \frac{T}{B}$

$\beta_u^2 = 1 - \frac{1}{(1+u')^2} \quad$ and $\quad u' = \frac{U}{mc^2} \quad$ and $\quad u = \frac{U}{B}$

$\beta_b^2 = 1 - \frac{1}{(1+b')^2} \quad$ and $\quad b' = \frac{B}{mc^2}$

where $\alpha, m, a_0$ and $c$ are the fine structure constant, the electron mass, the Bohr's radius and the speed of light in vacuum, respectively.

$B$ is the electron binding energy, $U$ the bound electron kinetic energy and $N$ the occupation number of the subshell to be ionized.

The relativistic form of the Vriens function $\phi$ is written as

$$\phi = \cos\left[\sqrt{\frac{\alpha^2}{(\beta_t^2 + \beta_b^2)}} Ln\left(\frac{\beta_t^2}{\beta_b^2}\right)\right] \quad \text{Eq 3}$$

In the BEB formalism of Kim *et al.* [34], the Vriens function is equal to its asymptotic form (=1).

The total cross section per shell $\sigma_{ion,MO}$ can be obtained from the single differential cross section (Eq 2) integrated from $w=0$ to $w=(t-1)/2$. The theoretical expression of this cross section per molecular orbital as a function of the incident energy is

$$\sigma_{ion,MO} = \frac{4\pi a_0^2 \alpha^4 N}{(\beta_t^2 + \beta_u^2 + \beta_b^2)2b'} \left[\left(1 - \frac{1}{t} + \frac{t-1}{2}\frac{b'^2}{(1+t'/2)^2}\right) - \phi\frac{Ln(t)}{t+1}\frac{1+2t'}{(1+t'/2)^2} \right. \quad \text{Eq 4}$$
$$\left. + \frac{1}{2}\left(Ln\left(\frac{\beta_t^2}{1-\beta_t^2}\right) - \beta_t^2 - Ln(2b')\right)\left(1 - \frac{1}{t^2}\right)\right]$$

The total ionisation cross section of the molecule is the sum of the cross sections (Eq 4) for all molecular orbitals:

$$\sigma_{ion}(T) = \sum_{1}^{nMO} \sigma_{ion,MO}(T) \quad \text{Eq 5}$$

The required data (B, U, and N) to calculate the cross sections for each MO of Eq2 and Eq4 are obtained from molecular electronic structure calculations since experimental data are not available. There are 35 MOs for adenine, 39 for guanine, 29 for cytosine and 33 for thymine. The number of inner shells is 10, 11, 8 and 9 for adenine, guanine, cytosine and thymine, respectively.

The molecular orbitals for the four DNA bases were obtained through calculations at restricted Hartree-Fock (RHF) level with basis-set cc-pVTZ on geometries previously optimized at the same level of theory. It is worth noting that basis-set cc-pVTZ is a correlation-consistent triple-zeta basis-set with polarization functions and, particularly, with Gausain-type functions up to the f shell. The corresponding data (B, U and N) were calculated as follows: B corresponds to the orbital energy of each MO resulting from the self-consistent resolution of the Hartree-Fock equations, U is the one-electron kinetic energy expectation value associated with the considered molecular orbital, and N is the molecular orbital occupation number, which, for an RHF computation on a 2n-electron closed-shell system, it is always equal to two for the first (occupied) n MOs and equal to zero for the remaining M-n (virtual) orbitals (with M as the number of basis functions used in the calculation). All the above-mentioned RHF computations were performed exploiting the quantum chemistry packages Gaussian09[35] and GAMESS-UK [36].

### 2.1.3 Excitation

Due to the lack of data for this process in DNA bases, the hypothesis to derive the electronic excitation cross section is based on the inelastic cross sections in water calculated by Dingfelder *et al.* [37]. In brief, the total excitation cross section is extracted from the total ionisation cross section of the base and the water models for excitation and ionisation (Eq 6), as proposed in the original version of the CPA100 code [23]. The assumption is that the ratio of the total ionisation cross section over total excitation cross section is the same in water and DNA components for each incident energy. For energies higher than 400 eV, this ratio in water tends to a constant.

$$\sigma_{exc,base} = \sigma_{ion,base} \left[\frac{\sigma_{exc}}{\sigma_{ion}}\right]_{water} \quad \text{Eq 6}$$

Only the electronic levels with a threshold lower than 20 eV can be excited in analogy with water [23]. With this assumption, there are 14, 12, 15 and 14 levels for adenine, cytosine, guanine and thymine, respectively, that can be excited.

The final hypothesis concerns how to select the levels. Without information, the probability is chosen to be the same for each level.

### 2.2 Implementation in Geant4-DNA

The electron interaction processes, elastic scattering, ionisation, and electronic excitation, were implemented as classes of Geant4-DNA physics models inherited from the G4VEmModel class [9] with Geant4.10.05.p01 version. Total and differential cross section data tables were calculated for each process as described in the previous sections. Each electron is tracked step by step and according to its energy prior to collision with the target material. A random process and a random target energy level are sampled according to the total cross section tables. For elastic process, the angle is randomly sampled according to the angular differential cross section tables and, for ionisation, the transferred energy is randomly sampled according to the energy differential cross section tables. Interpolation between consecutive energy points in the cross sections data tables was applied. Electrons are tracked down to 11 eV below which the track is killed and the energy is deposited locally. Therefore, the classes allow to simulate the stochastic nature of the electronic interactions.

The implementation of the new Geant4-DNA models for the four DNA bases was validated by performing three different calculations over the incident electron energies ranging from 11 eV to 1 MeV. Stopping power, range and inelastic mean free path of electrons were calculated using Geant4-DNA examples (called "spower", "range" and "mfp", respectively) developed for calculations in homogeneous targets of liquid water [12]. The targets in the three examples were replaced by uniform targets of adenine (A), thymine (T), guanine (G) and cytosine (C), respectively, and the physics models of water were replaced by new DNA bases models. Table 1 summarizes the model classes created for this study compared to liquid water models.

*Table 1 : Electron interaction processes, the physics models and the model classes used in the G4EmDNAPhysics_option6 constructor. The liquid water models are already available in the current public version of Geant4-DNA [22] and the four DNA bases models are presented in this work. Incident electron energy ranges are presented for each model.*

| | G4EmDNAPhysics_option6 | | |
|---|---|---|---|
| Process | Geant4-DNA model class | Liquid water models [22] | Four DNA bases models |
| Elastic Scattering | G4DNACPA100ElasticModel | Independent Atom Method model (11 eV – 256 keV) | Independent Atom Method model (11 eV – 1 MeV) |
| Ionisation | G4DNACPA100IonisationModel | Binary Encounter Bethe model (11 eV – 256 keV) | Relativistic Binary Encounter Bethe Vriens model (11 eV – 1 MeV) |
| Excitation | G4DNACPA100ExcitationModel | Dielectric model (11 eV – 256 keV) | Derived from water and ion cross sections (Eq 6) (11 eV- 1MeV) |

## 3. Results

To the best of our knowledge, we tried to collect all existing published data whether calculated or from experiments concerning DNA bases. In the following sections, the results of the quantities calculated in this study are shown in comparison with those published in the literature.

### 3.1 Electron interaction physics models: comparison with liquid water and available data
#### 3.1.1 Elastic scattering

Figure 1 presents two electron incident energies (100 eV and 10 keV) variation of the differential cross sections as a function of the scattering angle for the four DNA bases molecules. The results are compared with differential elastic cross section in water (water-option6, implemented in Geant4-DNA [22]), obtained from CPA100 code. For all incident energies and scattering angles, the elastic cross section in water is always lower than in DNA bases by about one order of magnitude as shown in Figure 1. The angular dependence of the cross sections values for the four bases is very similar. The lowest values of the differential cross sections are obtained for the smallest molecule (i.e. cytosine) and the highest for the largest (i.e. guanine) since the mass of the molecule, and the number of electrons are in the following decreasing order $G > A > T > C$. The observed minimum at 100 eV for all the molecules disappears at higher energies, showing a monotonical decrease as a function of the increasing scattering angle. The four bases exhibit the same minimum differential cross section value at 90° scattering angle.

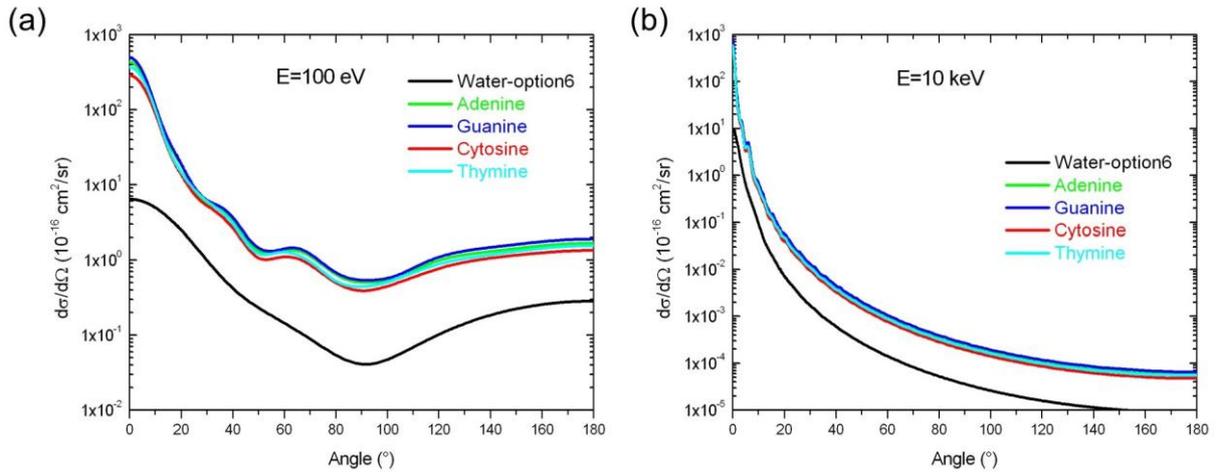

*Figure 1 : Elastic differential cross section for 100 eV (a) and 10 keV (b) electron collisions with adenine, guanine, cytosine, thymine and water respectively*

 

The integrated elastic cross sections of all four DNA bases and the liquid water option6 of Geant4-DNA are shown in Figure 2. The integrated elastic cross sections monotonically decrease with increasing incident energy (Figure 2). The integrated cross section curves of the DNA bases follow the same trend as the differential cross section curves where the molecular size dependence is observed ($\sigma_C < \sigma_T < \sigma_A < \sigma_G$, $\sigma$ stands for integrated cross section). As for differential cross section, the integrated elastic cross section in water is one order of magnitude lower than in the DNA bases.

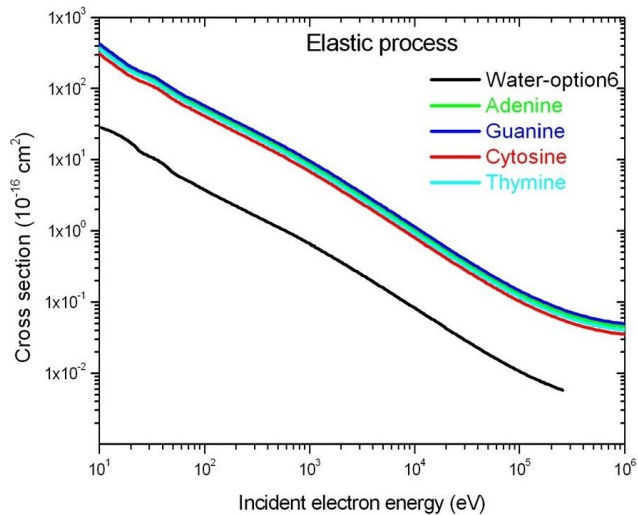

*Figure 2 : Integrated elastic cross section for adenine, guanine, cytosine and thymine and water collisions with electrons.*

Figure 3 shows as an example the elastic differential cross sections of 100 eV electrons for all four DNA bases in comparison with published calculations *[38-41]* of differential cross section curves.

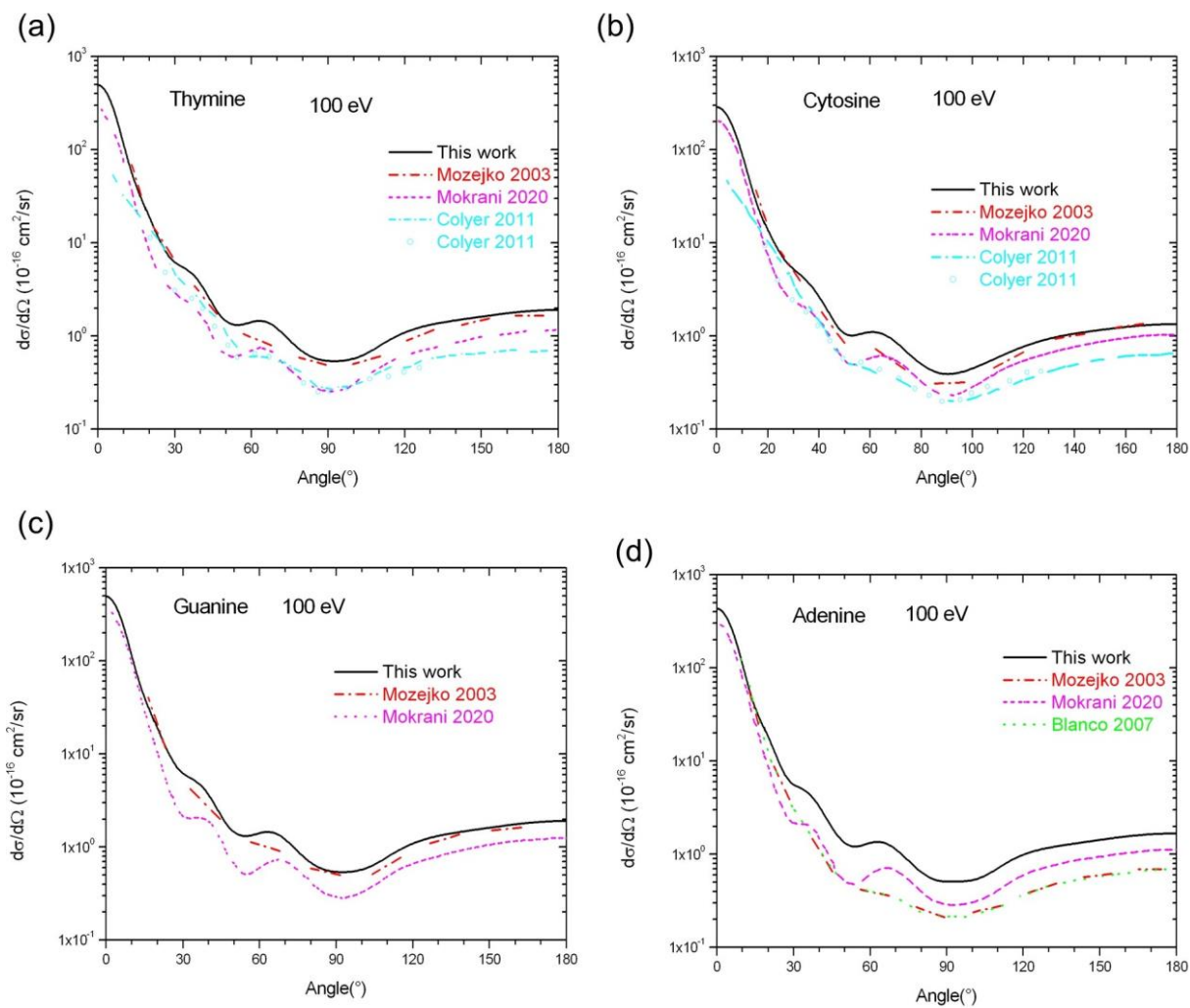

*Figure 3 : Elastic differential cross section for 100 eV incident electron energy from thymine (a), cytosine (b), guanine (c) and adenine (d), compared with published data (symbols are measurements [38] and lines are calculations [38-41]).*

Figure 4 presents a comparison of the integral elastic cross sections with available calculated data [39-44] obtained with approximation procedures for each base over a large incident energy range. The *ab initio* calculations limited to the low energy regime (between 0 and 20 eV for cytosine and thymine [45]) are not included.

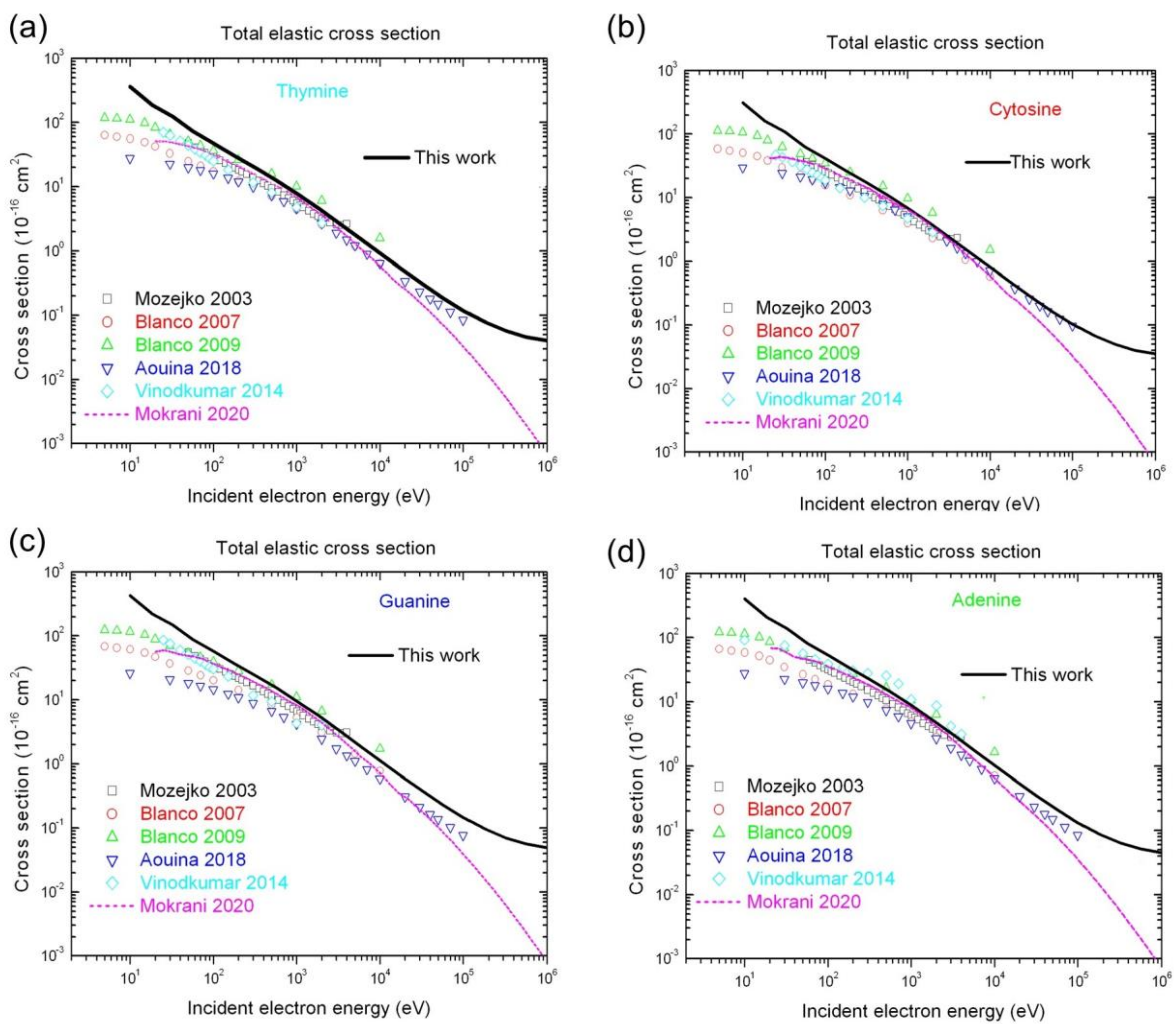

*Figure 4 : Elastic integrated cross section of electron collisions with thymine (a), cytosine (b), guanine (c) and adenine (d), compared with published calculations [39-44]*

### 3.1.2 Ionisation

The total ionisation cross section is obtained by summing the contribution of each orbital of the molecule calculated with eq 4. For each DNA base, Figure 5 gives the cross section for the sum of the valence shells (25 for adenine, 28 for guanine, 21 for cytosine and 24 for thymine) and the sum of the inner shells as a function of the incident energy between the threshold and 1 MeV. As we have observed, BEB calculations are more sensitive to binding energy than to mean kinetic energy.

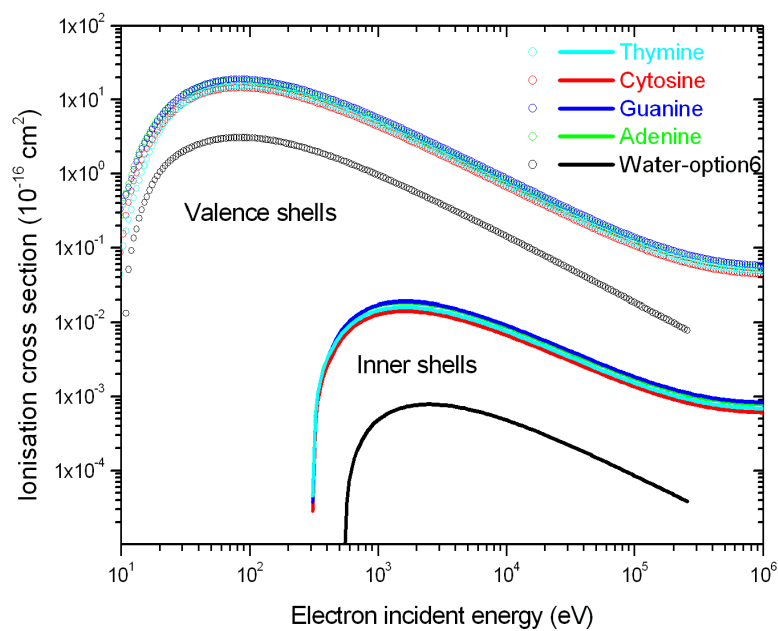

*Figure 5: The sum of valence shell ionisation cross sections (in lines), and the sum of internal shells ionisation cross sections (in symbols) for adenine, cytosine, guanine, thymine and water.*

In Figure 6 the total ionisation cross sections for the DNA bases and water option6 are shown.

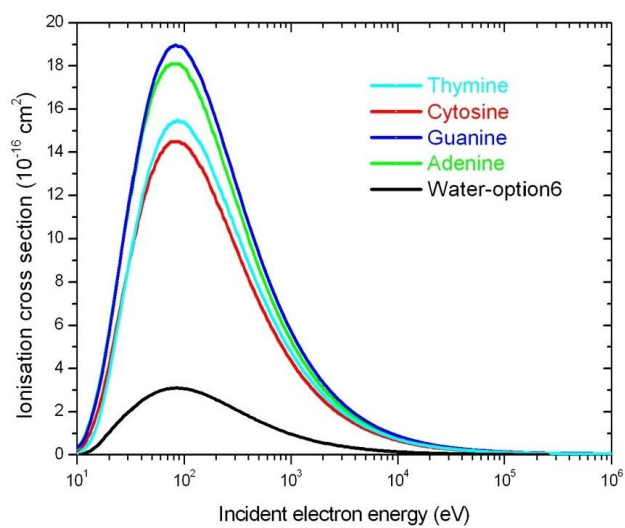

*Figure 6 : Total ionisation cross sections for the DNA bases and water.*

Figure 7 presents a comparison of experimental data [46-55], and theoretical calculations of the total ionisation cross section [39, 56-60] for the four bases. The theoretical method used is indicated in the legend of the figures.

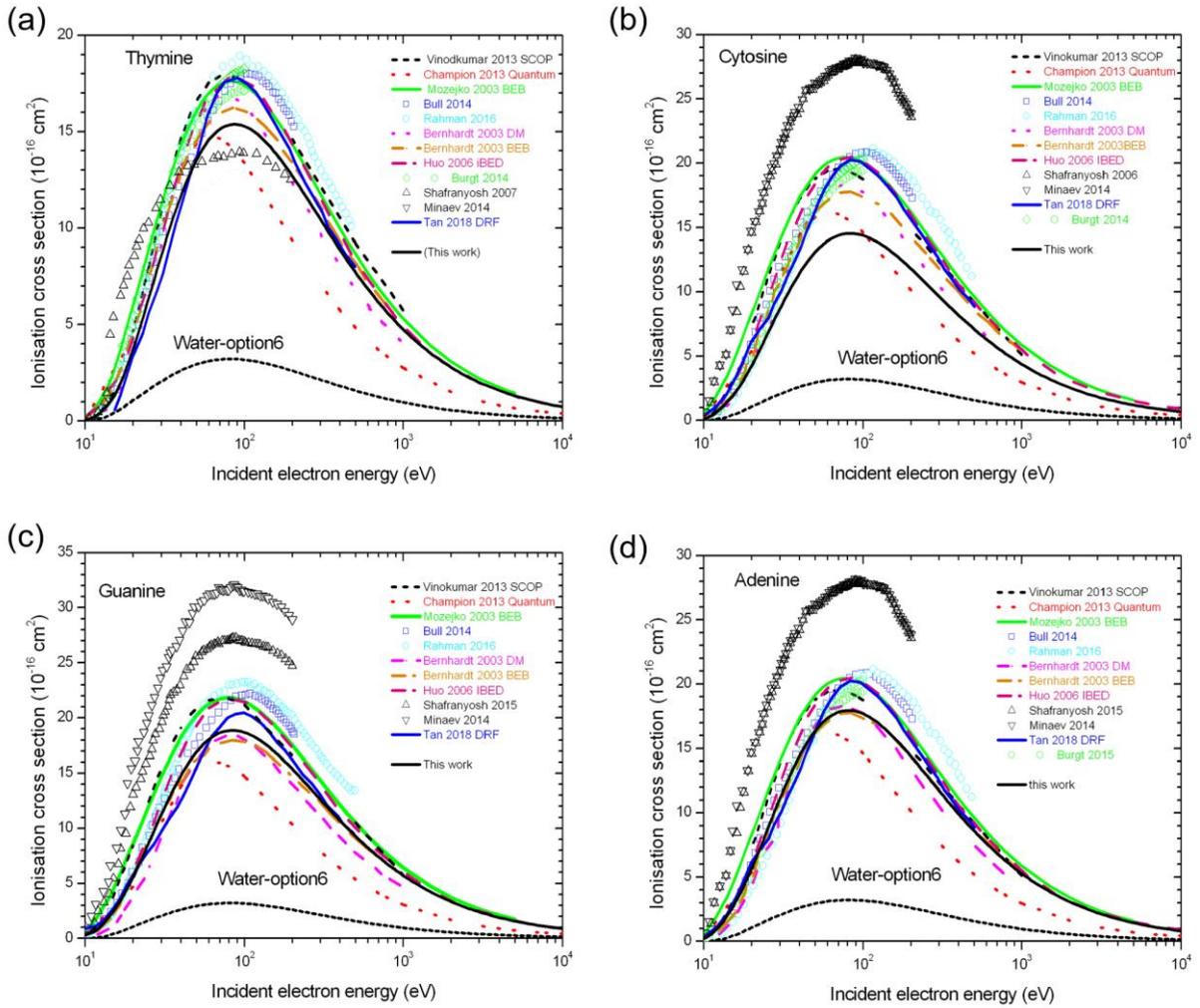

*Figure 7 : Total ionisation cross sections by electrons for the DNA bases, thymine (a), cytosine (b), guanine (c) and adenine (d), compared with available data (lines: calculations and the used method [39, 56-60], symbols and symbols: measurements [46-55] ). For comparison, the cross section in water is added.*

The energy differential cross sections (EDCS) were calculated using Eq. 2. Figure 8.a represents the variation of the EDCS for thymine at different incident energies as a function of the electron ejected energy. The results are compared to CPA100 data for water previously calculated in Geant4-DNA water-option6 [22]. The differential cross section in water is always lower than in thymine for any incident energy. The same trend is observed for other DNA bases (not plotted). If we compare the variation of the EDCS for one incident electron energy for the four DNA bases and water, the trend is similar to what was obtained for the elastic and total ionisation cross sections: the impact of the size of the molecule has the same effect on the variations of the EDCS (Figure 8.b at 1 keV). Analogous results are obtained with other incident energies (same curve shape, not shown here). There are no experimental data with which to compare these calculations.

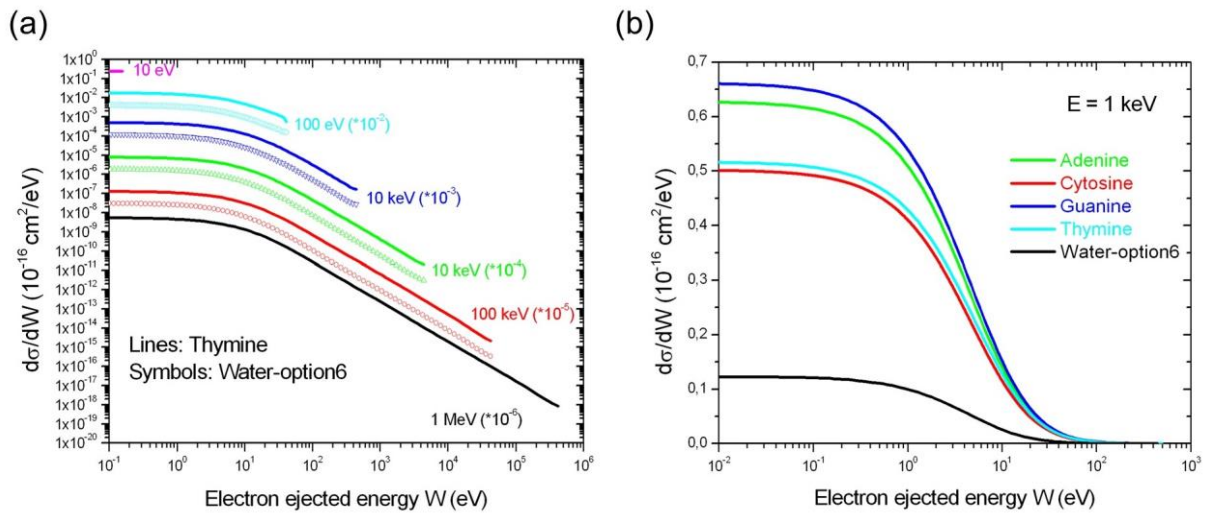

*Figure 8 : Differential cross sections by electrons as a function of the electron ejected energy in thymine and in water at different incident electron energies(a), and in the four bases and water at 1keV incident electron energy (b).*

### 3.1.3 Electronic excitation

Figure 9 presents the calculated total electronic excitation cross section for the four DNA bases obtained using eq 6. Except at low energy (<≈30 eV), the variations are directly related to the molecular size, as observed for the total ionisation cross sections. Being an inelastic process, the maximum appears for energy lower than 100 eV, at approximately the same value regardless of the DNA base. Like other processes, the cross section for electronic excitation in water is lower than in the DNA bases.

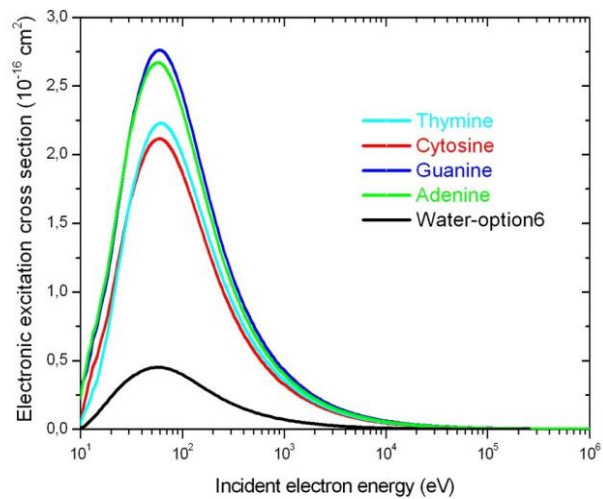

*Figure 9 : Electronic excitation collision cross section in the DNA compounds.*

## 3.2 Verification of physics models in Geant4-DNA

### 3.2.1 Stopping power

Stopping power was calculated for the four DNA bases and compared with theoretical calculations [61-65], previous Geant4-DNA Monte Carlo [20] simulations and data derived from the input material data file of PENELOPE code, as shown in figure 10 and were also compared with calculations in water-option6. The values were compared with calculations of Akar *et al.*

[63, 64] and ESTAR [61] database. Calculations by Akkerman *et al.* [65] and Joy [62] are only available for guanine.

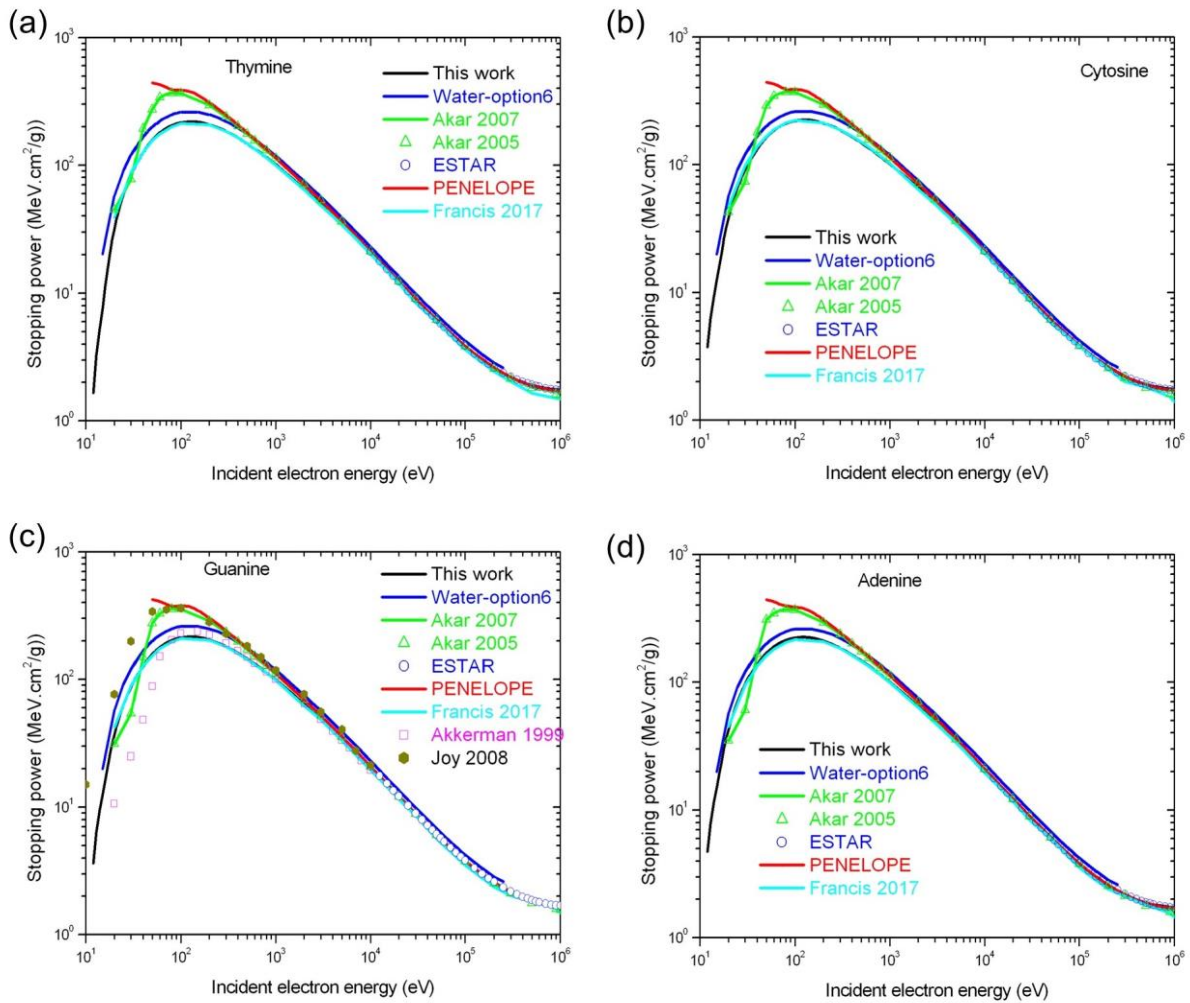

*Figure 10: Stopping power of the four nucleobases for electrons ranging from 11 eV - 1 MeV as calculated in this work in comparison with Geant4-DNA water-option6 simulation and other results from the literature [20, 61-65]. Stopping power in the four nucleobases was derived from the input material data file of PENELOPE code.*

### 3.2.2 Range

The range is calculated for all four bases and compared with results from the literature and water-option6 Geant4-DNA simulation as Figure 11 shows. Calculations by Akar *et al.* [63, 64], Akkerman *et al.* [65] (guanine only) and ESTAR [61] are shown. Values derived from PENELOPE code are also plotted for comparison.

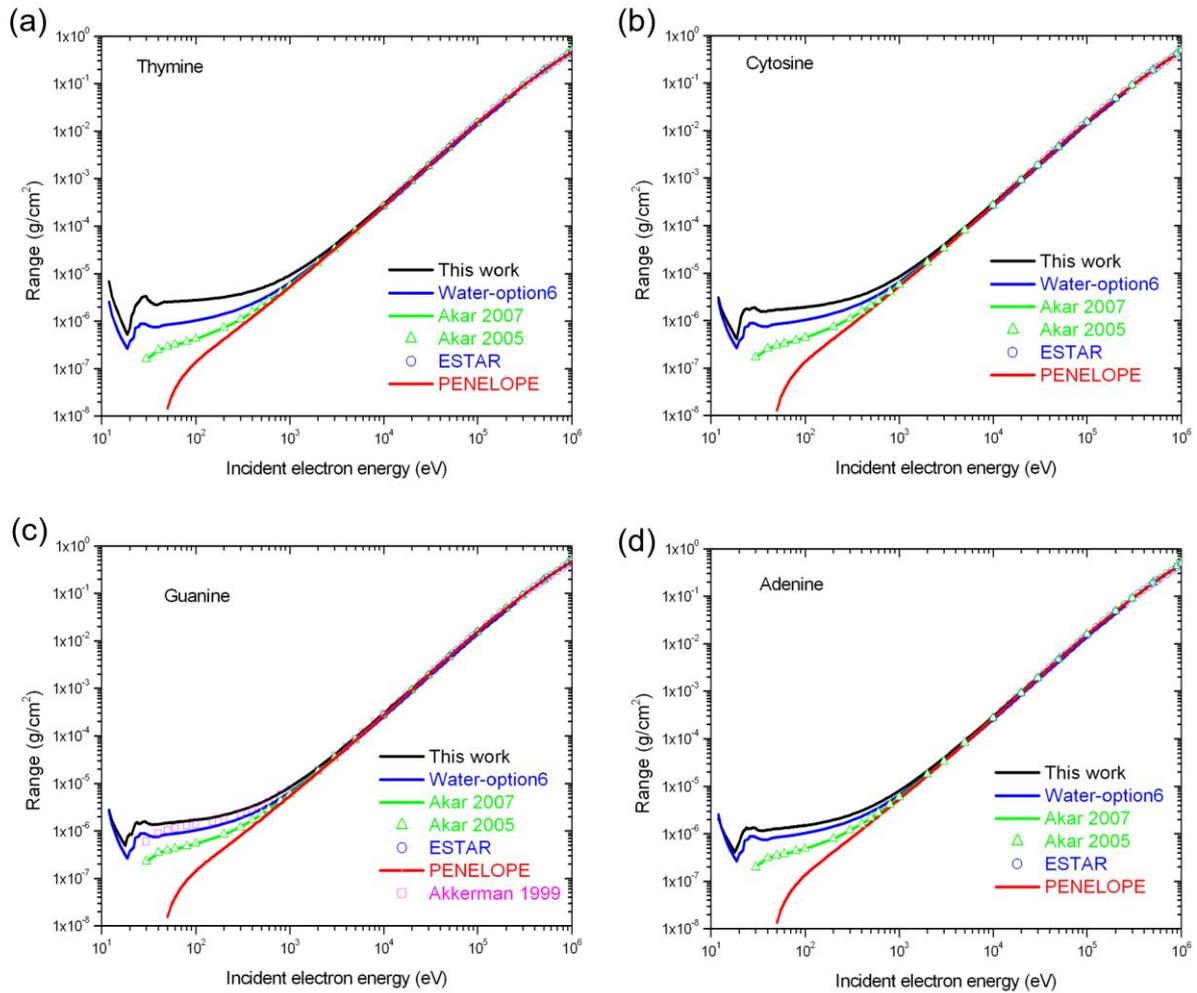

*Figure 11: Range of electrons ranging from 11 eV - 1 MeV in the four nucleobases as calculated in this work in comparison with Geant4-DNA water-option6 simulation and other results from the literature [61, 63-65]. Range in the four nucleobases was derived from the input material data file of PENELOPE code.*

### 3.2.3  Inelastic mean free path

In Figure 12, the IMFP of electrons in the four DNA bases and water are shown. Comparison with calculations of Akar *et al.* [66] and Tanuma *et al.* [67] (for adenine and guanine) are also shown.

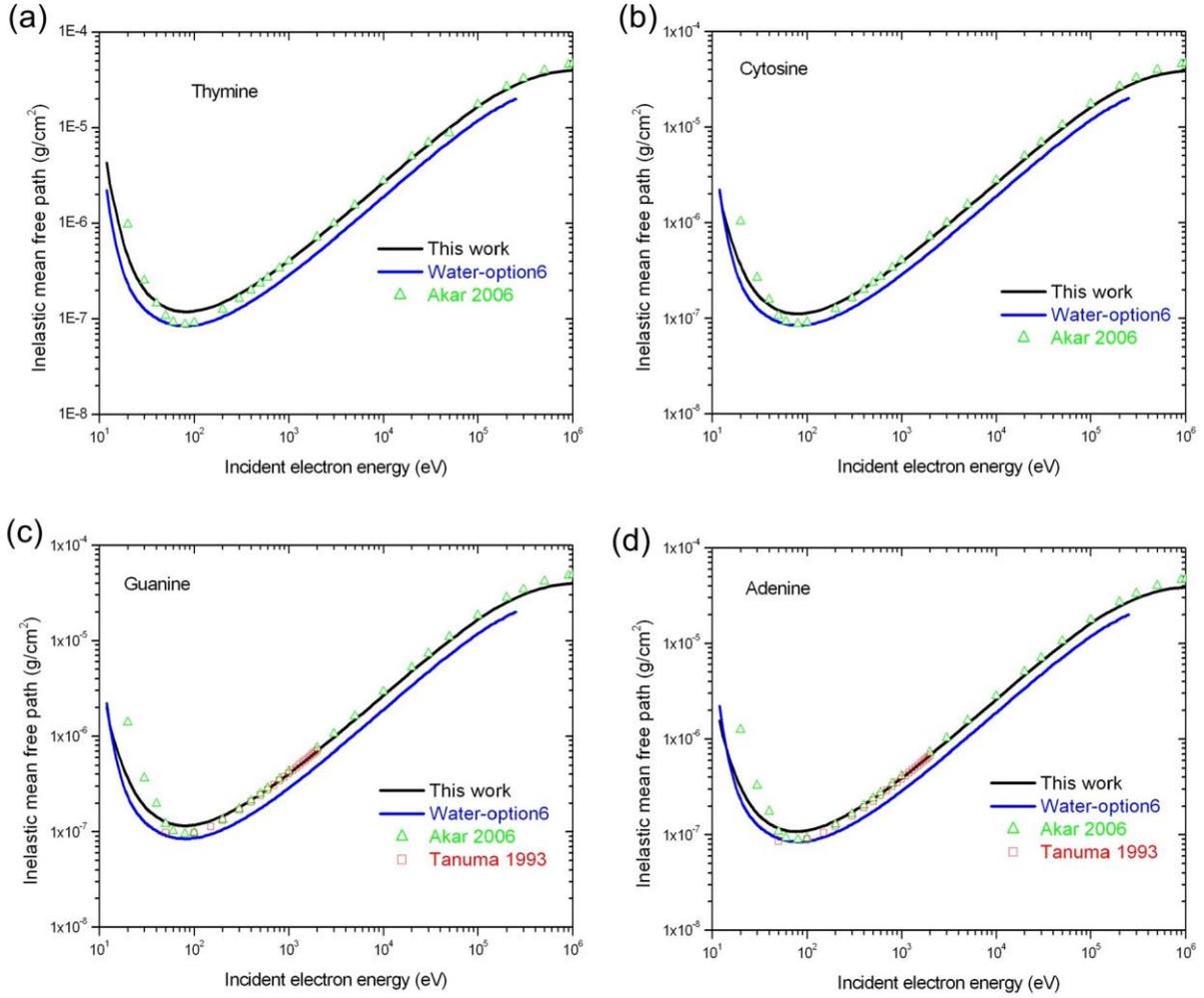

*Figure 12: Inelastic mean free path of electrons ranging from 11 eV - 1 MeV in the four nucleobases as calculated in this work in comparison with Geant4-DNA water-option6 simulation and other results from the literature [66, 67]. IMFPs from Tanuma et al. [67] were normalized to the nucleobase densities as reported by Tan et al. [68]*

## 4. Discussion

There is a direct target size dependence of the calculated electron cross sections. Obviously larger molecules induce higher interaction probability that reflects into the elastic and inelastic total and differential cross sections. This clear distinction between the different DNA bases and the water targets emphasises the importance of tracking electrons in the actual biomolecules. Since the cross sections of biomolecules are larger than those of water, the simulation with water alone will underestimate the damage.

There are few studies that calculated elastic differential cross sections for DNA bases and no experimental data is available apart from the single study by Colyer *et al.* [38] where only thymine and cytosine differential cross sections were measured for six discrete incident electron energies between 60 eV and 500 eV and scattering angles between 15° and 130°. Other studies calculated theoretical differential cross sections for the four bases [38-41] using different approximations. Colyer *et al.* [38] provided theoretical values for cytosine and thymine, and Blanco *et al.*[41] provided theoretical values for adenine only. Apart from the study of Mokrani et al., which calculated cross sections in the 10 eV – 5 MeV[40], these studies are limited to low energy ranges (50 – 4000 eV[39] and 5 – 10000 eV[41]). The differential cross sections calculated in this work are in good agreement with data from the literature regarding shape and magnitude, as shown in Figure 3. Oscillations (Figure 3 for 100 eV incident energy) are clearly observed in the recent calculations of Mokrani *et al.* [40], although the values are lower than around a factor of 2. The observed discrepancies can be attributed to the different potentials used in the calculations.

All integral elastic cross section results of Figure 4 are in a relatively good agreement at the intermediate energy range (200 eV-10 keV). At low energy (< 100 eV), our results are systematically higher. All the available integral elastic cross sections data take into account different potentials and partial wave expansions in their calculations. The values obtained by Blanco and Garcia [42] are based on a simplified procedure of the method used in their previous paper [41]. The differences observed at high energy with the results obtained by Mokrani *et al.* [40] are due to the fact that they do not introduce correction terms linked with relativistic effects. Aouina and Chaoui [43] used the same screen-corrected additivity rule as Blanco and Garcia [41], but with a different decomposition based on the relativistic Dirac partial wave expansion instead of the Schrödinger one. The method proposed by Vinodkumar *et al.* [44] is based on the Schrödinger equation with spherical potential for molecules including exchange and polarization.

For all molecules, a peak is observed around 80 eV in the total ionisation cross section (Figure 6a). Since the ionisation potentials are approximately the same, the peaks are very close to each other. Similarly to elastic scattering and except at very low energy (< 25 eV) where the influence of the ionisation potential is important, the amplitude of total ionisation cross section is linked to the size of the molecule: the highest cross section is for guanine, which is the largest molecular system investigated in this work. The total ionisation cross section in water is around one order of magnitude lower than in the DNA bases. If we compare the ionisation cross section per valence electron (which is the sum of the cross section for all the levels divided by the number of valence electron of the base), the differences between water and DNA bases vanish because both molecules are built with low-Z atoms which are covalently bonded.

The total ionisation cross sections (Figure 7) exhibit the same typical shape with a maximum – whose amplitude varies according to the model used – and a gradual decrease towards high impact energies. The results are scarce and the height and the corresponding energy of the peak differ. However, the same quantitative behaviour of curves is observed and differences mainly appear in the magnitude mostly expressed at the peaks.

The limited number of experimental results [46-55] is mainly due to the preparation requirements of these pure target-biomolecules in the gas phase. The experiments carried out between 2006 and 2015, are limited to low energy (<500 eV). Experimental data [50, 52-55] are generally far from the theoretical results.

The theoretical studies devoted to ionisation impact in DNA bases are mainly based on semi-classical formalisms. An evident shift regarding the position of the maximum predicted by different models may be noted. Some calculations are based on BEB formalism or derived from it [39, 56, 58]. The oldest ones [39, 56] used low-order Hartree-Fock method and smaller basis set in the geometry optimization and the differences in the results are due to the target description and more particularly the vertical ionisation potential. In the work of Huo *et al.* [58], the correlation-consistent polarized valence double-zeta (cc-pVDZ) basis set of Gaussian functions was used. The results are comparable to the cross sections of Mozejko and Sanche [39], but are larger than those of Bernhardt and Paretzke [56]. Even with the same formalism, there are variations in the results linked to the parameters used. For example, the choice of the binding energy of the first ionisation potential differs depending on whether it is obtained from experiments or calculations and induces differences in the cross section especially at low energy. That is why main discrepancies with BEB results appear at low binding energy.

Bernhardt and Paretzke [56] calculated the total ionisation cross sections using two formalisms: BEB and Deutsch-Märk (DM). Both formalisms show quite a similar trend; however, DM calculations reach higher peaks at slightly higher energy than BEB and the decrease in the DM curve is steeper compared to the BEB one.

Other types of calculation [57, 59, 60] cover a larger energy range, although limited to 10 keV. Vinodkumar *et al.* [60] used the spherical complex optical potential formalism to calculate the inelastic cross section, and the group additivity rule. For their results, the best agreement is found with the DM results of Bernhardt and Paretzke [56] in adenine and cytosine. In guanine, the best agreement is observed with the results obtained by Huo *et al.* [58] and with the DM results beyond the peak. For thymine, the best agreement is obtained with Mozejko and Sanche [39] and Huo *et al.* [58]. Recently, in 2013, Champion [57] developed a theoretical quantum mechanical model to calculate the differential and the total ionisation cross sections of the different DNA components up to 10 keV. In this case, the curve rises more rapidly than the other results. Tan *et al.* [59] used a simple semi-empirical method based on dielectric formalism for proton interaction extended by adding the exchange interaction of electrons and by taking into account a mean binding energy to correctly reproduce the low energy part of the cross section.

Our results are comparable to the other available results in amplitude and location of the peak. More precisely, the position of the peak in our calculations is in better agreement with the position of the calculations of dielectric response function (DRF) [59], DM [56], improved Binary Encounter Dipole (iBED) [58], although the amplitude is lower. The difference from the ionisation total cross section in water is important.

ESTAR [61] stopping powers, calculated from the theory of Bethe are only available for high incident electron energies (>10 keV) and our results are in good agreement with less than 4% average difference at 10 keV, decreasing to 1% at 1 MeV for all four DNA bases (Figure 11). A general agreement in the shape of the stopping power curves is observed for the different calculations. The peak stopping power was at 80 eV in the calculations of Akar *et al.* [63, 64], and at 100 eV in the Monte Carlo simulations of Francis *et al.* [20] similar to our results, whereas the calculations performed by Akkerman *et al.* [65] reached the peak stopping power at 120 eV. At low incident energies < 1keV, theoretical calculations of Akar *et al.* [63, 64] using Bethe theory with Generalized Oscillator Strength method score higher stopping power somewhat similar to the results of Joy [62]. The inelastic stopping power [63] and total stopping power including Bremsstrahlung effects [64] calculated by the two studies of Akar *et al.* are almost identical for low energy electrons. For energies higher than 300 keV, Bremsstrahlung effects are observed with 1% difference between both stopping powers increasing to about 6.4% for 1 MeV electrons.

The best agreement for our results is with those obtained by Francis *et al.* [20] for all four DNA bases, which used the Rudd model cross sections implementation in Geant4-DNA. When compared with water-option6 calculations, the stopping power of all four DNA bases is consistently lower over the whole incident energy range.

Good agreement is observed with range results by ESTAR [61] for all four bases, with a 10% difference at 10 keV incident electron energy decreasing to 1% at 1MeV (Figure 12). Our results are in good agreement with calculations of Akar *et al.* [63, 64] and Akkerman *et al.* [65] (guanine only) as well as PENELOPE data for energies higher than 1 keV. However, for lower energies a deviation is observed between all types of calculations since each of them follows a different physical model. Electrons have smaller range in water compared to DNA bases and this reflects the higher stopping power of water, as shown in Figure 10.

There are not plenty of results in the literature for IMFP. Therefore, in Figure 12 we show a comparison with calculations done by Akar *et al*. [66] for the four DNA bases and with those carried out by Tanuma *et al.* [67] for adenine and guanine only over a short incident electron energy range (50-2000 eV). IMFPs from Tanuma *et al.* [67] were normalized to the nucleobase densities as reported by Tan *et al.* [68] The IMFP plots (Figure 12) show a reasonably good agreement with the results of Akar *et al.* and of Tanuma *et al.* with small deviations at low energies (< 100 eV). As expected, an inverse relation with stopping power is shown and electrons in water score consistently lower values than in the DNA bases.

In this study, the various electron cross sections with the four DNA bases were calculated using distinct physics models. To guarantee the correct implementation of these cross sections in Geant4-DNA, it was important to verify our simulation results with published data for available energy ranges, even though the physical models used in the literature may differ from ours. The good agreement of our calculations with published data shows the reliability of the physical models and the simulations. The stopping power, range and IMFP calculations show obvious difference between the DNA bases and the water targets as well. The obvious distinction between water results and biomolecules therefore requires transport codes capable of tracking particles in various biomolecules, which is introduced in this work.

## 5. Conclusion

There is a growing need for the availability of track structure Monte Carlo simulations in DNA material to model the detailed particle-biological matter interactions. This study introduced a new set of electron interaction cross sections in the four DNA nucleobases over a large incident electron energy range. The cross sections were successfully implemented in the open source Geant4-DNA simulation toolkit. Good agreement with experiments and calculations from the literature in terms of cross sections, stopping power, range and inelastic mean free path was shown. The results of this study will enable the extension of Geant4-DNA classes to be used in various biological molecules in addition to liquid water targets. The obvious differences of interaction cross sections and physical parameters between DNA bases and water will induce differences in energy deposition and their localization, which will affect the radiation damage estimation and, therefore, the importance of accurate simulation of these data.

**Acknowledgments**

The authors would like to thank Professor F. Salvat for kindly providing the most recent version of ELSEPA used in the work. S. Zein and S. Incerti thank IN2P3/CNRS for the funding support to Geant4-DNA.

**References**

[1] D.R. White, J. Booz, R.V. Griffith, J.J. Spokas, I.J. Wilson, Report 44, Journal of the International Commission on Radiation Units and Measurements, os23 (2016) NP-NP.
[2] I. El Naqa, P. Pater, J. Seuntjens, Monte Carlo role in radiobiological modelling of radiotherapy outcomes, Physics in Medicine and Biology, 57 (2012) R75-97.
[3] M. Terrissol, A. Beaudre, Simulation of space and time evolution of radiolytic species induced by electrons in water, Radiation Protection Dosimetry, 31 (1990) 175-177.

[4] H. Nikjoo, S. Uehara, D. Emfietzoglou, F.A. Cucinotta, Track-structure codes in radiation research, Radiation Measurements, 41 (2006) 1052-1074.
[5] A. Peudon, S. Edel, M. Terrissol, Molecular basic data calculation for radiation transport in chromatin, Radiation protection dosimetry, 122 (2006) 128-135.
[6] W. Friedland, E. Schmitt, P. Kundrát, M. Dingfelder, G. Baiocco, S. Barbieri, A. Ottolenghi, Comprehensive track-structure based evaluation of DNA damage by light ions from radiotherapy-relevant energies down to stopping, Scientific reports, 7 (2017) 45161.
[7] J. Baro, J. Sempau, J. Fernández-Varea, F. Salvat, PENELOPE: an algorithm for Monte Carlo simulation of the penetration and energy loss of electrons and positrons in matter, Nuclear Instruments and Methods in Physics Research Section B: Beam Interactions with Materials and Atoms, 100 (1995) 31-46.
[8] J.M. Fernández-Varea, G. González-Muñoz, M.E. Galassi, K. Wiklund, B.K. Lind, A. Ahnesjö, N. Tilly, Limitations (and merits) of PENELOPE as a track-structure code, International journal of Radiation Biology, 88 (2012) 66-70.
[9] S. Incerti, G. Baldacchino, M. Bernal, R. Capra, C. Champion, Z. Francis, P. Guèye, A. Mantero, B. Mascialino, P. Moretto, P. Nieminen, C. Villagrasa, C. Zacharatou, THE GEANT4-DNA PROJECT, International Journal of Modeling, Simulation, and Scientific Computing, 01 (2010) 157-178.
[10] S. Incerti, A. Ivanchenko, M. Karamitros, A. Mantero, P. Moretto, H.N. Tran, B. Mascialino, C. Champion, V.N. Ivanchenko, M.A. Bernal, Z. Francis, C. Villagrasa, G. Baldacchino, P. Guèye, R. Capra, P. Nieminen, C. Zacharatou, Comparison of GEANT4 very low energy cross section models with experimental data in water, Med Phys, 37 (2010) 4692-4708.
[11] M.A. Bernal, M.C. Bordage, J.M.C. Brown, M. Davídková, E. Delage, Z. El Bitar, S.A. Enger, Z. Francis, S. Guatelli, V.N. Ivanchenko, M. Karamitros, I. Kyriakou, L. Maigne, S. Meylan, K. Murakami, S. Okada, H. Payno, Y. Perrot, I. Petrovic, Q.T. Pham, A. Ristic-Fira, T. Sasaki, V. Štěpán, H.N. Tran, C. Villagrasa, S. Incerti, Track structure modeling in liquid water: A review of the Geant4-DNA very low energy extension of the Geant4 Monte Carlo simulation toolkit, Physica Medica: European Journal of Medical Physics, 31 (2015) 861-874.
[12] S. Incerti, I. Kyriakou, M.A. Bernal, M.C. Bordage, Z. Francis, S. Guatelli, V. Ivanchenko, M. Karamitros, N. Lampe, S.B. Lee, S. Meylan, C.H. Min, W.G. Shin, P. Nieminen, D. Sakata, N. Tang, C. Villagrasa, H.N. Tran, J.M.C. Brown, Geant4-DNA example applications for track structure simulations in liquid water: A report from the Geant4-DNA Project, Med Phys, 45 (2018) e722-e739.
[13] S. Agostinelli, J. Allison, K. Amako, J. Apostolakis, H. Araujo, P. Arce, M. Asai, D. Axen, S. Banerjee, G. Barrand, GEANT4—a simulation toolkit, Nucl Inst Meth Phys Res A, 506 (2003) 250-303.
[14] J. Allison, K. Amako, J. Apostolakis, H. Araujo, P.A. Dubois, M. Asai, G. Barrand, R. Capra, S. Chauvie, R. Chytracek, G.A.P. Cirrone, G. Cooperman, G. Cosmo, G. Cuttone, G.G. Daquino, M. Donszelmann, M. Dressel, G. Folger, F. Foppiano, J. Generowicz, V. Grichine, S. Guatelli, P. Gumplinger, A. Heikkinen, I. Hrivnacova, A. Howard, S. Incerti, V. Ivanchenko, T. Johnson, F. Jones, T. Koi, R. Kokoulin, M. Kossov, H. Kurashige, V. Lara, S. Larsson, F. Lei, O. Link, F. Longo, M. Maire, A. Mantero, B. Mascialino, I. McLaren, P.M. Lorenzo, K. Minamimoto, K. Murakami, P. Nieminen, L. Pandola, S. Parlati, L. Peralta, J. Perl, A. Pfeiffer, M.G. Pia, A. Ribon, P. Rodrigues, G. Russo, S. Sadilov, G. Santin, T. Sasaki, D. Smith, N. Starkov, S. Tanaka, E. Tcherniaev, B. Tome, A. Trindade, P. Truscott, L. Urban, M. Verderi, A. Walkden, J.P. Wellisch, D.C. Williams, D. Wright, H. Yoshida, Geant4 developments and applications, IEEE Transactions on Nuclear Science, 53 (2006) 270-278.
[15] J. Allison, K. Amako, J. Apostolakis, P. Arce, M. Asai, T. Aso, E. Bagli, A. Bagulya, S. Banerjee, G. Barrand, B.R. Beck, A.G. Bogdanov, D. Brandt, J.M.C. Brown, H. Burkhardt, P. Canal, D. Cano-Ott, S. Chauvie, K. Cho, G.A.P. Cirrone, G. Cooperman, M.A. Cortés-Giraldo, G. Cosmo, G. Cuttone, G. Depaola, L. Desorgher, X. Dong, A. Dotti, V.D. Elvira, G. Folger, Z. Francis, A. Galoyan, L. Garnier, M. Gayer, K.L. Genser, V.M. Grichine, S. Guatelli, P. Guèye, P. Gumplinger, A.S. Howard, I. Hřivnáčová, S. Hwang, S. Incerti, A. Ivanchenko, V.N. Ivanchenko, F.W. Jones, S.Y. Jun, P. Kaitaniemi, N. Karakatsanis, M. Karamitros, M. Kelsey, A. Kimura, T. Koi, H. Kurashige, A. Lechner, S.B. Lee, F. Longo, M. Maire, D. Mancusi, A. Mantero, E. Mendoza, B. Morgan, K. Murakami, T. Nikitina, L.

Pandola, P. Paprocki, J. Perl, I. Petrović, M.G. Pia, W. Pokorski, J.M. Quesada, M. Raine, M.A. Reis, A. Ribon, A. Ristić Fira, F. Romano, G. Russo, G. Santin, T. Sasaki, D. Sawkey, J.I. Shin, I.I. Strakovsky, A. Taborda, S. Tanaka, B. Tomé, T. Toshito, H.N. Tran, P.R. Truscott, L. Urban, V. Uzhinsky, J.M. Verbeke, M. Verderi, B.L. Wendt, H. Wenzel, D.H. Wright, D.M. Wright, T. Yamashita, J. Yarba, H. Yoshida, Recent developments in Geant4, Nucl Inst Meth Phys Res A, 835 (2016) 186-225.

[16] M. Karamitros, A. Mantero, S. Incerti, W. Friedland, G. Baldacchino, P. Barberet, M. Bernal, R. Capra, C. Champion, Z. ElBitar, Z. Francis, P. Guèye, A. Ivanchenko, V. Ivanchenko, H. Kurashige, B. Mascialino, P. Moretto, P. Nieminen, G. Santen, H. Seznec, H. Tran, C. Villagrasa, C. Zacharatou, Modeling Radiation Chemistry in the Geant4 Toolkit, Progress in Nuclear Science and Technology, 2 (2011) 503-508.

[17] W. Friedland, P. Jacob, P. Bernhardt, H.G. Paretzke, M. Dingfelder, Simulation of DNA damage after proton irradiation, Radiation research, 159 (2003) 401-410.

[18] Z. Francis, C. Villagrasa, I. Clairand, Simulation of DNA damage clustering after proton irradiation using an adapted DBSCAN algorithm, Computer Methods and Programs in Biomedicine, (2011) 265-270.

[19] J.D. Watson, F.H.C. Crick, Molecular Structure of Nucleic Acids: A Structure for Deoxyribose Nucleic Acid, Nature, 171 (1953) 737-738.

[20] Z. Francis, Z.E. Bitar, S. Incerti, M.A. Bernal, M. Karamitros, H.N. Tran, Calculation of lineal energies for water and DNA bases using the Rudd model cross sections integrated within the Geant4-DNA processes, Journal of Applied Physics, 122 (2017) 014701.

[21] M.U. Bug, W. Yong Baek, H. Rabus, C. Villagrasa, S. Meylan, A.B. Rosenfeld, An electron-impact cross section data set (10eV–1keV) of DNA constituents based on consistent experimental data: A requisite for Monte Carlo simulations, Radiation Physics and Chemistry, 130 (2017) 459-479.

[22] M.-C. Bordage, J. Bordes, S. Edel, M. Terrissol, X. Franceries, M. Bardies, N. Lampe, S. Incerti, Implementation of new physics models for low energy electrons in liquid water in Geant4-DNA, Physica Medica, 32 (2016) 1833-1840.

[23] S. Edel, Modélisation du transport des photons et des électrons dans l'ADN plasmide, Université Toulouse III-Paul Sabatier, Toulouse France, 2006.

[24] A. Peudon, Prise en compte de la structure moléculaire pour la modélisation des dommages biologiques radio-induits, Université Toulouse III - Paul Sabatier, 2007.

[25] N.F. Mott, H.S.W. Massey, The theory of atomic collisions, Oxford Clarendon Press 1965.

[26] Chemical Structures Project, 2009.

[27] F. Salvat, A. Jablonski, C.J. Powell, ELSEPA—Dirac partial-wave calculation of elastic scattering of electrons and positrons by atoms, positive ions and molecules, Computer physics communications, 165 (2005) 157-190.

[28] D. Sakata, S. Incerti, M.-C. Bordage, N. Lampe, S. Okada, D. Emfietzoglou, I. Kyriakou, K. Murakami, T. Sasaki, H. Tran, An implementation of discrete electron transport models for gold in the Geant4 simulation toolkit, Journal of Applied Physics, 120 (2016) 244901.

[29] W.-G. Shin, M.-C. Bordage, D. Emfietzoglou, I. Kyriakou, D. Sakata, C. Min, S.B. Lee, S. Guatelli, S. Incerti, Development of a new Geant4-DNA electron elastic scattering model for liquid-phase water using the ELSEPA code, Journal of Applied Physics, 124 (2018) 224901.

[30] Y.-K. Kim, M.E. Rudd, Binary-encounter-dipole model for electron-impact ionization, Physical Review A, 50 (1994) 3954-3967.

[31] Y.-K. Kim, W. Hwang, N.M. Weinberger, M.A. Ali, M.E. Rudd, Electron-impact ionization cross sections of atmospheric molecules, J Chem Phys, 106 (1997) 1026-1033.

[32] I. Torres, R. Martínez, M.N.S. Rayo, F. Castaño, Evaluation of the computational methods for electron-impact total ionization cross sections: Fluoromethanes as benchmarks, The Journal of chemical physics, 115 (2001) 4041-4050.

[33] M. Guerra, P. Amaro, J. Machado, J.P. Santos, Single differential electron impact ionization cross sections in the binary-encounter-Bethe approximation for the low binding energy regime, Journal of Physics B: Atomic, Molecular and Optical Physics, 48 (2015) 185202.

[34] Y.-K. Kim, J.P. Santos, F. Parente, Extension of the binary-encounter-dipole model to relativistic incident electrons, Physical Review A, 62 (2000) 052710.

[35] M. Frisch, G. Trucks, H. Schlegel, G. Scuseria, M. Robb, J. Cheeseman, G. Scalmani, V. Barone, B. Mennucci, G. Petersson, H. Nakatsuji, M. Caricato, X. Li, H. Hratchian, A. Izmaylov, J. Bloino, G. Zheng, J. Sonnenberg, M. Hada, M. Ehara, K. Toyota, R. Fukuda, J. Hasegawa, M. Ishida, T. Nakajima, Y. Honda, O. Kitao, H. Nakai, T. Vreven, J.J. Montgomery, J. Peralta, F. Ogliaro, M. Bearpark, J. Heyd, E. Brothers, K. Kudin, V. Staroverov, R. Kobayashi, J. Normand, K. Raghavachari, A. Rendell, J. Burant, S. Iyengar, J. Tomasi, M. Cossi, N. Rega, J. Millam, M. Klene, J. Knox, J. Cross, V. Bakken, C. Adamo, J. Jaramillo, R. Gomperts, R. Stratmann, O. Yazyev, A. Austin, R. Cammi, C. Pomelli, J. Ochterski, R. Martin, K. Morokuma, V. Zakrzewski, G. Voth, P. Salvador, J. Dannenberg, S. Dapprich, A. Daniels, Ö. Farkas, J. Foresman, J. Ortiz, J. Cioslowski, D. Fox, Gaussian 09, Gaussian Inc., Wallingford, CT, USA, Revision D.01 (2009).

[36] M.F. Guest, I.J. Bush, H.J.J. Van Dam, P. Sherwood, J.M.H. Thomas, J.H. Van Lenthe, R.W.A. Havenith, J. Kendrick, The GAMESS-UK electronic structure package: algorithms, developments and applications, Molecular Physics, 103 (2005) 719-747.

[37] M. Dingfelder, D. Hantke, M. Inokuti, H.G. Paretzke, Electron inelastic-scattering cross sections in liquid water, Radiat Phys Chem, 53 (1998) 1-18.

[38] C. Colyer, S. Bellm, F. Blanco, G. García, B. Lohmann, Elastic electron scattering from the DNA bases cytosine and thymine, Physical Review A, 84 (2011) 042707.

[39] P. Możejko, L. Sanche, Cross section calculations for electron scattering from DNA and RNA bases, Radiation and Environmental Biophysics, 42 (2003) 201-211.

[40] S. Mokrani, H. Aouchiche, C. Champion, Elastic scattering of electrons by DNA bases, Radiation Physics and Chemistry, 172 (2020) 108735.

[41] F. Blanco, G. García, Calculated cross sections for electron elastic and inelastic scattering from DNA and RNA bases, Physics Letters A, 360 (2007) 707-712.

[42] F. Blanco, G. García, A screening-corrected additivity rule for the calculation of electron scattering from macro-molecules, Journal of Physics B: Atomic, Molecular and Optical Physics, 42 (2009) 145203.

[43] N.Y. Aouina, Z.-E.-A. Chaoui, Simulation of positron and electron elastic mean free path and diffusion angle on DNA nucleobases from 10 eV to 100 keV, Surface and Interface Analysis, 50 (2018) 939-946.

[44] M. Vinodkumar, C. Limbachiya, H. Desai, P.C. Vinodkumar, Electron-impact total cross sections for phosphorous triflouride, Physical Review A, 89 (2014) 062715.

[45] C. Winstead, V. McKoy, S. d'Almeida Sanchez, Interaction of low-energy electrons with the pyrimidine bases and nucleosides of DNA, The Journal of chemical physics, 127 (2007) 085105.

[46] J.N. Bull, J.W. Lee, C. Vallance, Absolute electron total ionization cross-sections: molecular analogues of DNA and RNA nucleobase and sugar constituents, Physical chemistry chemical physics : PCCP, 16 (2014) 10743-10752.

[47] P.J.M. van der Burgt, Electron impact fragmentation of cytosine: partial ionization cross sections for positive fragments, The European Physical Journal D, 68 (2014) 135.

[48] P.J.M. van der Burgt, F. Mahon, G. Barrett, M.L. Gradziel, Electron impact fragmentation of thymine: partial ionization cross sections for positive fragments, The European Physical Journal D, 68 (2014) 151.

[49] P.J.M. van der Burgt, S. Finnegan, S. Eden, Electron impact fragmentation of adenine: partial ionization cross sections for positive fragments, The European Physical Journal D, 69 (2015) 173.

[50] B. Minaev, M. Shafranyosh, Y. Svida, M. Sukhoviya, I. Shafranyosh, G. Baryshnikov, V. Minaeva, Fragmentation of the adenine and guanine molecules induced by electron collisions, The Journal of chemical physics, 140 (2014) 175101.

[51] M.A. Rahman, E. Krishnakumar, Communication: Electron ionization of DNA bases, The Journal of chemical physics, 144 (2016) 161102.

[52] I.I. Shafranyosh, M.I. Sukhoviya, M.I. Shafranyosh, Absolute cross sections of positive- and negative-ion production in electron collision with cytosine molecules, Journal of Physics B: Atomic, Molecular and Optical Physics, 39 (2006) 4155-4162.

[53] I. Shafranyosh, M. Sukhoviya, Electron impact excitation of gas-phase thymine molecules, Optics and Spectroscopy, 102 (2007) 500-502.

[54] I. Shafranyosh, M. Sukhoviya, M. Shafranyosh, L. Shimon, Formation of positive and negative ions of thymine molecules under the action of slow electrons, Technical Physics, 53 (2008) 1536-1540.

[55] I. Shafranyosh, Y.Y. Svida, M. Sukhoviya, M. Shafranyosh, B. Minaev, G. Baryshnikov, V. Minaeva, Absolute effective cross sections of ionization of adenine and guanine molecules by electron impact, Technical Physics, 60 (2015) 1430-1436.

[56] P. Bernhardt, H.G. Paretzke, Calculation of electron impact ionization cross sections of DNA using the Deutsch–Märk and Binary–Encounter–Bethe formalisms, International Journal of Mass Spectrometry, 223-224 (2003) 599-611.

[57] C. Champion, Quantum-mechanical predictions of electron-induced ionization cross sections of DNA components, The Journal of chemical physics, 138 (2013) 184306.

[58] W.M. Huo, C.E. Dateo, G.D. Fletcher, Molecular data for a biochemical model of DNA damage: Electron impact ionization and dissociative ionization cross sections of DNA bases and sugar-phosphate backbone, Radiation measurements, 41 (2006) 1202-1208.

[59] H.Q. Tan, Z. Mi, A.A. Bettiol, Simple and universal model for electron-impact ionization of complex biomolecules, Physical review. E, 97 (2018) 032403.

[60] M. Vinodkumar, C. Limbachiya, M.Y. Barot, M. Swadia, A. Barot, Electron impact total ionization cross sections for all the components of DNA and RNA molecule, International Journal of Mass Spectrometry, 339– 340 (2013) 16– 23.

[61] M.J. Berger, J.S. Coursey, M.A. Zucker, J. Chang, ESTAR, PSTAR, and ASTAR: Computer Programs for Calculating Stopping-Power and Range Tables for Electrons, Protons, and Helium Ions (version 1.2.3). National Institute of Standards and Technology, Gaithersburg, MD., 2005.

[62] D.C. Joy, A Database of Electron-Solid Interactions, 2008.

[63] A. Akar, H. Gümüş, Electron stopping power in biological compounds for low and intermediate energies with the generalized oscillator strength (GOS) model, Radiation Physics and Chemistry, 73 (2005) 196-203.

[64] A. Akar, H. Gümüş, N. Okumuşoğlu, Total electron stopping powers and CSDA-ranges from 20 eV to 10 MeV electron energies for components of DNA and RNA, Advances in Quantum Chemistry, 52 (2007) 277-288.

[65] A. Akkerman, E. Akkerman, Characteristics of electron inelastic interactions in organic compounds and water over the energy range 20–10000 eV, Journal of Applied Physics, 86 (1999) 5809-5816.

[66] A. Akar, H. Gümüş, N.T. Okumusoglu, Electron inelastic mean free path formula and CSDA-range calculation in biological compounds for low and intermediate energies, Applied radiation and isotopes : including data, instrumentation and methods for use in agriculture, industry and medicine, 64 (2006) 543-550.

[67] S. Tanuma, C.J. Powell, D.R. Penn, Calculations of electron inelastic mean free paths (IMFPS). IV. Evaluation of calculated IMFPs and of the predictive IMFP formula TPP-2 for electron energies between 50 and 2000 eV, Surface and Interface Analysis, 20 (1993) 77-89.

[68] Z. Tan, Y. Xia, M. Zhao, X. Liu, Proton stopping power in a group of bioorganic compounds over the energy range of 0.05–10MeV, Nuclear Instruments and Methods in Physics Research Section B: Beam Interactions with Materials and Atoms, 248 (2006) 1-6.